\documentstyle[12pt]{article}
\begin{document}

\input{epsf}

\title{QED correction to asymmetry for
polarized $ep$--scattering from the method of the electron structure
functions}

\author{A.V. Afanasev$^{a)}$, I. Akushevich$^{a)}$, N.P.Merenkov$^{b)}$ }

\date{}
\maketitle
\begin{center}
{\small {\it $^{(a)}$ North Carolina Central University,
Durham, NC 27707, USA \\ and  \\
TJNAF, Newport News, VA 23606, USA\\}}
{\small {\it{$^{(b)}$ NSC "Kharkov Institute of Physics and Technology" \\
}}}
{\small {\it {61108, Akademicheskaya 1, Kharkov, Ukraine}}}
\end{center}
\begin{abstract}
The electron structure function method is applied to calculate
model--independent radiative corrections to an asymmetry of
electron--proton scattering.  The representations for both
spin--independent and spin--dependent parts of the cross--section are
derived.  Master formulae take into account the leading corrections
in all orders and the main contribution of the second order
next--to--leading ones and have accuracy at the level of one per mille.
Numerical calculations illustrate our analytical results for both
elastic and deep inelastic events.
\end{abstract}

\section{Introduction}
\hspace{0.6cm}

Precise polarization measurements in both inclusive \cite{JF,Anse} and
elastic \cite{JLabel,NIKIEF} scattering are crucial for understanding
the structure and fundamental properties of a nucleon.


One important component of the precise data analysis is radiative
effects, which always accompany the processes of electron
scattering. The  first calculation of radiative corrections (RC) to
polarized deep inelastic
scattering (DIS) was done
by Kukhto and Shumeiko \cite{KS},  who applied a covariant method of
extraction of an infrared divergence \cite{BS,bardinrev} to this
process.
The polarization states were described by 4-vectors, which were kept in
their general forms during the calculation.  It required tedious
procedure of tensor integration over photonic phase space, and, as a
result, led to very complicated structure of final formulae for RC.
The next step was done in the paper \cite{ASh}, where additional
covariant
expansion of polarization 4-vectors over a certain basis allowed to
simplify
the calculation and final results. It resulted in producing the
Fortran code POLRAD \cite{POLRAD} and Monte Carlo generator RADGEN
\cite{Radgen}.	These tools are widely used in all
current experiments in polarized DIS.  Later the calculation was applied
to the case of collider experiments on DIS \cite{Bard,AIShJP}.
We applied
this
method also to elastic process in papers \cite{AAM,AAIM}.

However, the method of covariant extraction of infrared divergence is
essentially restricted by the lowest order RC.	All attempts to go beyond
the lowest order lead to very large formulae, that are
difficult to cross check, or to a simple leading log approach
\cite{AIShappr}. The recent developments are reviewed
in ref.\cite{echaja}.

The decision can be found in applying the formalism of the electron
structure functions (ESF).  Within this approach such processes as the
electron--positron annihilation into hadrons and the deep inelastic
electron--proton scattering in one photon exchange approximation can be
considered as the Drell--Yan process \cite{DY} in annihilation or
scattering channel, respectively.  Therefore, the QED radiative
corrections (RC) to the corresponding cross--sections can be written as
a contraction of two electron structure functions and the hard part of
the cross--section, see \cite{KF,KMF}.
  Traditionally these RC include effects caused by loop
  corrections as well as soft and hard collinear radiation of
  photons and $e^+e^-$-pairs. But it
was shown in Ref. \cite{KMF}
how one
can
  improve this method by inclusion also effects due to radiation
  of one non-collinear photon. The corresponding procedure
  concludes in
  modification of the hard part of cross section that provides
  the exact accounting of the lowest order correction and
  leads to exit beyond the leading approximation.
We applied this approach to the
recoil proton polarization in elastic electron scattering in
ref.\cite{AAMrec}.  In the present paper we calculate RC to polarized
DIS and elastic scattering following  ref.\cite{AAMrec}.

Section \ref{sectESF} gives a short introduction to the structure
function method. There we present two known forms of the electron
structure functions, namely,  iterative and analytical, which resums
singular
  infrared terms in all order into exponent.
 In this section we also obtain
master formulae for observed cross sections. Leading log results
are presented in Section \ref{sectLL}. These results are valid both for
DIS and elastic cases. We also use an iterative form of ESF to extract
the lowest order correction, that can provide a cross-check through
comparison
with known results. In Sections \ref{sectNLODIS} and \ref{sectNLOel} we
describe the procedure of generalization the results
 for next-to-leading order in DIS and elastic cases.
Numerical analysis is presented in Section \ref{sectNum}. We consider
kinematical conditions of current polarization experiments at fixed
targets as well as collider kinematics. Some conclusions are made in the
Section \ref{sectConc}.

\section{Electron Structure Functions}\label{sectESF}

A straightforward calculation based on the quasireal electron method
\cite{BFK} can be used to write the invariant cross--section of the
DIS
process
\begin{equation}\label{1}
e^-(k_1) + P(p_1) \rightarrow e^-(k_2) + X(p_x) \
\end{equation}
in the following form
\begin{equation}\label{2,master formula}
\frac{d\sigma(k_1,k_2)}{d\ Q^2\,d\ y} =
\int\limits_{z_{1m}}^1d\,z_1\int \limits_{z_{2m}}^1d\,z_2
D(z_1,L)\frac{1}{z_2^2} D(z_2,L)
\frac{d^2\sigma_{hard}(\tilde k_1,\tilde k_2)}{d\tilde Q^2\,d\tilde y} \
, \ \ L=\ln\frac{Q^2}{m^2}\ ,
\end{equation} where $m$ is the electron mass
and $$Q^2=-(k_1-k_2)^2, \ y=\frac{2p_1(k_1-k_2)}{V}\ , \ \ V=2p_1k_1
.$$

The reduced variables which define the hard cross--section in the
integrand are
\begin{equation}\label{3, tilde variables}
\tilde k_1 = z_1k_1\ , \ \ \tilde k_2 = \frac{k_2}{z_2}\ , \ \
\widetilde Q^2 = \frac{z_1}{z_2}Q^2\ , \ \
\tilde y = 1 -\frac{1-y}{z_1z_2}\ .
\end{equation}

The electron structure function $D(z,L)$ includes contributions due to
photon emission and pair production
\begin{equation}\label{4, unpol.  EST}
D = D^{^{\gamma}} + D^{^{e^+e^-}}_N + D^{^{e^+e^-}}_S \ ,
\end{equation}
where $D^{^{\gamma}}$ is responsible for the photons radiation and
$D^{^{e^+e^-}}_N $ and $D^{^{e^+e^-}}_S$ describe pair production in
nonsinglet (by single photon mechanism) and singlet (by double
photon mechanism) channels, respectively.

The structure functions on the right-hand side of Eq.~(4) satisfy the
DGLAP equations \cite{LAP} (see also \cite{KF}). The
functions $D(z_1,L)$ and $D(z_2,L)$ is responsible for radiation of the
initial and final electrons, respectively.

There exist different representations for the photonic contribution
into the structure function \cite{KF,JSW,S} but here we will use the
form given in \cite{KF} for $D^{\gamma}$, $D^{^{e^+e^-}}_N$ and
$D^{^{e^+e^-}}_S$
\begin{equation}\label{5}
D^{\gamma}(z,Q^2) =
\frac{1}{2}\beta(1-z)^{\beta/2-1}\Bigl[1+\frac{3}{8}\beta
-\frac{\beta^2}{48}\bigl(\frac{1}{3}L+\pi^2-\frac{47}{8}\bigr)\Bigr]-\frac
{\beta}{4}(1+z) +
\end{equation}
$$\frac{\beta^2}{32}\Big[-4(1+z)\ln(1-z)-\frac{1+3z^2}{1-z}\ln\,z-5-z\Bigr]
\ , \ \ \beta = \frac{2\alpha}{\pi}(L-1)\ . $$
\begin{equation}\label{6,pair contribution into EST}
D^{^{e^+e^-}}_N(z,Q^2) =
\frac{\alpha^2}{\pi^2}\Bigl[\frac{1}{12(1-z)}\bigl(1-z-\frac{2m}{\varepsilon}
\bigr)^{\beta/2}\bigl(L_1-\frac{5}{3}\bigr)^2\bigl(1+z^2+\frac{\beta}{6}
\bigl(L_1-\frac{5}{3}\bigr)\bigr)\Bigr]
\theta\bigl(1-z-\frac{2m}{\varepsilon}\bigr) \ ,
\end{equation}
\begin{equation}\label{7}
D^{^{e^+e^-}}_S = \frac{\alpha^2}{4\pi^2}L^2\bigl[\frac{2(1-z^3)}{3z}
+\frac{1}{2}(1-z)
+(1+z)\ln{z}\bigr]\theta\bigl(1-z-\frac{2m}{\varepsilon}\bigr) \ .
\end{equation}
where $\varepsilon$ is the energy of the parent electron and $L_1= L +
2\ln(1-z).$
Note that the above form of the structure function $D^{^{e^+e^-}}_N$
includes effects due to real pair production only. The correction caused
by the virtual pair is included in $D^{^{\gamma}}$. Terms containing
contribution
of the order $\alpha^2L^3$ are cancelled out in the sum $D^{^{\gamma}} +
D^{^{e^+e^-}}_N.$

Instead of the photon structure function given by Eqs.~(5)--(7), one can
use their iterative form \cite{JSW} \begin{equation}\label{8,
iterative form EST} D^{\gamma}(z,L) = \delta(1-z) +
\sum_{k=1}^{\infty}\frac{1}{k!}
\biggl(\frac{\alpha\,L}{2\pi}\biggr)^kP_1(z)^{\otimes k}\ ,
\end{equation}
$$\underbrace{ P_1(z)\otimes\cdots\otimes P_1(z)}_{k} = P_1(z)^{\otimes k}
\ , \quad P_1(z)\otimes P_1(z) = \int\limits_{
z}^{1}P_1(t)P_1\biggl(\frac{z}{t} \biggr)\frac{dt}{t} \ ,$$
$$P_1(z) = \frac{1+z^2}{1-z}\theta(1-z-\Delta) +
\delta(1-z)\bigl(2\ln{\Delta}+\frac{3}{2}\bigr) \ , \ \Delta \ll 1 \ . $$
The iterative form (8) of $D^{^{\gamma}}$ does not include any
effects caused by pair production. The corresponding
nonsinglet part of the structure due to real and virtual pair production
can be included into the iterative form of
$D^{\gamma}(z,L)$ by replacing $\alpha L/2\pi$ on the right-hand
side of Eq.~(8) with the effective electromagnetic coupling
\begin{equation}\label{9,effective coupling}
\frac{\alpha L}{2\pi}\rightarrow\frac{\alpha_{eff}}{2\pi} =
-\frac{3}{2}\ln{\bigl(1-\frac{\alpha L}{3\pi}\bigr)} \end{equation} that
is (within the leading accuracy) the integral of the running
electromagnetic constant.

The lower limits of integration with respect to $z_1$ and $z_2$ in the
master
Eq.(2) can be obtained from the condition for existence of inelastic
hadronic
events
\begin{equation}\label{10}
(p_1+\tilde q)^2>M_{th}^2 \ , \ \ \tilde q = \tilde k_1 -\tilde k_2\ ,
\ M_{th} = M+m_{\pi}\ ,
\end{equation}
where $m_{\pi}$ is the pion mass. This constraint can be rewritten in
terms of dimensionless variables as follows
\begin{equation}\label{11, constraint}
z_1z_2+y-1-xyz_1\leq z_2z_{th}\ , \ \ x = \frac{Q^2}{2p_1(k_1-k_2)}\ , \ \
z_{th} = \frac{M_{th}^2-M^2}{V}\ ,
\end{equation}
which leads to
$$z_{2m}= \frac{1-y+xyz_1}{z_1-z_{th}}, \qquad
z_{1m}= \frac{1+z_{th}-y}{1-xy}
 .$$

The squared matrix element of the considered process in one photon
exchange approximation is proportional to contraction of the leptonic and
hadronic tensors. The representation (2) reflects the properties of the
leptonic tensor. Therefore, it has the universal nature (because of
universality of the leptonic tensor) and can be applied to processes with
different final hadronic states. In particular, we can use
the electron structure function method to compute RC to the elastic
and deep inelastic (inclusive and semi-inclusive) electron--proton
scattering cross--sections.

On the other hand, the straightforward calculations in the first order
with respect
to $\alpha$ \cite{BFK,KS,ASh} and the recent calculations of the leptonic
current tensor in the second order \cite{AAK,GKM,KM,KMS} for the
longitudinally polarized initial electron demonstrate that in the
leading approximation spin--dependent and spin--independent parts of
this tensor are the same for the nonsinglet channel contribution.  The
latter corresponds to photon radiation and $e^+e^-$--pair
production through the single--photon mechanism. The difference appears
in the
second order due to possibility of pair production in the singlet
channel by
double--photon mechanism \cite{KMS}. Therefore, the representation (2),
being slightly modified, can be used for the calculation of RC to
cross--sections of different processes with a longitudinally polarized
electron beam.

In our recent work \cite{AAMrec} we applied the electron structure
function
method to compute RC to the ratio of the recoil proton polarizations
measured at CEBAF by Jefferson Lab Hall A Collaboration \cite{JLabel}.
The aim of this high precision experiment is the measurement of the proton
electric formfactor $G_E.$ In the present work we use this method for
calculation of model--independent part of RC to the asymmetry in
scattering of
longitudinally polarized electrons on polarized protons at the level of per
mile accuracy for elastic and deep inelastic hadronic events.

The cross--section of the scattering of the longitudinally
polarized electron by the proton with given longitudinal $(\parallel)$ or
transverse $(\bot)$ polarization for both elastic and deep inelastic
events can be written as a sum of the spin--independent and
spin--dependent parts \begin{equation}\label{12}
\frac{d\sigma(k_1,k_2,S)}{d\,Q^2d\,y}=
\frac{d\sigma(k_1,k_2)}{d\,Q^2d\,y}+\eta\frac{d\sigma^{^{\parallel,\bot}}
(k_1,k_2,S)}{d\,Q^2d\,y}\ ,
\end{equation}
where $S$ is the 4--vector of the target proton polarization and
$\eta$
is
the product of the electron and proton polarization degrees. Herein
after
we assume $\eta =1.$

The master Eq.(2) describes the RC to the spin--independent part of the
cross--section on the right-hand side of Eq.~(12) and the corresponding
equation for the
spin--dependent part reads
\begin{equation}
\label{13,master formula for polarized case}
\frac{d\sigma{^{\parallel,\bot}}(k_1,k_2,S)}{d\ Q^2\,d\
y} = \int\limits_{z_{1m}}^1d\,z_1\int \limits_{z_{2m}}^1d\,z_2
D^{(p)}(z_1,L)\frac{1}{z_2^2} D(z_2,L)
\frac{d^2\sigma^{{\parallel,\bot}}_{hard}(\tilde k_1,\tilde
k_2,S)}{d\tilde Q^2\,d\tilde y} \ ,
\end{equation}
where
$$D^{(p)} = D^{\gamma}+ D^{^{e^+e^-}}_N + D^{^{e^+e^-}(p)}_S\ , $$
and \cite{KMS}
\begin{equation}\label{14}
D^{^{e^+e^-}(p)}_S=\frac{\alpha^2}{4\pi^2}L^2\bigl(\frac{5(1-z)}{2}+(1+z)
\ln\,z\bigr)\theta\bigl(1-z-\frac{2m}{\varepsilon}\bigr)\ ,
\end{equation}
describes the radiation of the initial polarized electron.

     The representation is valid if radiation of collinear
   particles does not lead to change of polarization 4--vectors
   $S^{\parallel}$ and $S^{\bot}.$ In general it is not so
   \cite{AAGM},
 but in
   this paper we use just such polarizations which satisfy this
   condition (see below Eq.(18)).

The asymmetry in elastic scattering and DIS processes is defined as a
ratio
\begin{equation}\label{15, asymmetry}
A^{^{\parallel,\bot}}=
\frac{d\sigma^{^{\parallel,\bot}}(k_1,k_2,S)}{d\sigma(k_1,k_2)}\ ,
\end{equation}
therefore RC to the asymmetry requires the knowledge of
RC to both spin--independent and spin--dependent parts of the
cross--section.

RC to the spin--independent part were calculated (within the
electron structure function approach) in \cite{KMF}. In the present work
we compute the RC to spin--dependent parts for longitudinal and
transverse polarizations of the target proton and longitudinally
polarized electron beam. To be complete, we repeat briefly the result for
unpolarized case.


\section{The leading approximation}
\label{sectLL}

\hspace{0.7cm}
Within the leading accuracy (by taking into account the terms of the
order
$(\alpha L)^n$ the electron structure function can be computed, in principle,
in all orders of the perturbation theory. In this approximation, we have to
take the Born cross--section as a hard part on the right-hand sides of Eqs.~(2)
and (13).

We express the Born cross--section in terms of leptonic and hadronic
tensors as follows
\begin{equation}\label{16, Born normalization}
\frac{d\sigma}{dQ^2\,dy}=
\frac{4\pi\alpha^2(Q^2)}{VQ^4}L_{\mu\nu}^B\,H_{\mu\nu}\ ,
\end{equation}
where $\alpha(Q^2)$ is the running electromagnetic constant that accounts for
the effects of vacuum polarization and
\begin{equation}\label{17, hadronic tensor}
H_{\mu\nu} = -F_1\widetilde g_{\mu\nu}+\frac{F_2}{p_1q}\tilde
p_{1\mu}\tilde p_{1\nu}
-i\frac{M\epsilon_{\mu\nu\lambda\rho}q_{\lambda}}{p_1q}
\bigl[(g_1+g_2)S_{\rho} - g_2\frac{Sq}{p_1q}p_{1\rho}\bigr] \ ,
\end{equation}
$$L_{\mu\nu}^B = -\frac{Q^2}{2}g_{\mu\nu}
+k_{1\mu}k_{2\nu}+k_{1\nu}k_{2\mu} + i\varepsilon_{\mu\nu\lambda\rho}
q_{\lambda}k_{1\rho}\  , \
\widetilde g_{\mu\nu}=g_{\mu\nu}- \frac{q_{\mu}q_{\nu}}{q^2}\ , \ \tilde
p_{1\mu} = p_{\mu}- \frac{p_1q}{q^2}q_{\mu} \ .$$ In Eqs.~(17)
we assume the proton and electron polarization degrees equal to $1$.
The spin--independent $(F_1,\ F_2)$ and spin--dependent $(g_1,\ g_2)$
proton structure functions depend on two variables
$$x'=\frac{-q^2}{2p_1q}\ , \ \ q^2 = (p_x-p_1)^2\ . $$ In Born
approximation $x'=x$, but they differ in general case, when radiation of
photons and electron--positron pairs is allowed.

It is convenient to parameterize the 4--vector of proton polarization in
the form \cite{ASh}
\begin{equation}\label{18, representation for polarization}
S^{\parallel}_{\mu} = \frac{2M^2k_{1\mu}-Vp_{1\mu}}{MV}\ , \
S^{\bot}_{\mu} =
\frac{up_{1\mu}+Vk_{2\mu}-[2u\tau+V(1-y)]k_{1\mu}}{\sqrt{-uV^2(1-y)-u^2M^2}}
\ ,  \end{equation}
where $u=-Q^2,$ $\tau = M^2/V.$

One can verify that the 4--vector $S^{^{\parallel}}$ in the laboratory
system has components $(0,\vec n),$ where 3--vector
$\vec n$ has orientation of the initial electron 3--momentum $\vec k_1.$
One can verify also that
$S^{^{\bot}}S^{^{\parallel}}=0$ and in the laboratory system
$$S^{^{\bot}} = (0,\vec n_{\bot})\ , \ \ \vec n_{\bot}^2 = 1\ ,  \ \ \vec
n \vec n_{\bot} =0 \ , $$ where 3--vector $\vec n_{\bot}$ belongs to the
plane $(\vec k_1, \vec k_2).$

As normalization is chosen, the elastic limit $(p_x^2 =M^2)$ can be
reached by a simple substitution in the hadronic tensor
\begin{equation}\label{19, elastic limit}
F_1(x',q^2)\rightarrow \frac{1}{2}\delta(1-x')G_M^2(q^2)\ , \ F_2(x',q^2)
\rightarrow \delta(1-x')\frac{G_E^2(q^2)+\lambda G_M^2(q^2)}{1+\lambda}\ ,
\end{equation}
$$g_1(x',q^2)\rightarrow\frac{1}{2}\delta(1-x')\bigl\{G_M(q^2)G_E(q^2)+
\frac{\lambda}{1+\lambda}[G_M(q^2)-G_E(q^2)]G_M(q^2)\bigr\} , $$
$$g_2(x',q^2)\rightarrow
-\frac{1}{2}\delta(1-x')\frac{\lambda}{1+\lambda}[G_M(q^2)-G_E(q^2)]G_M(q^2)\
, \
\lambda = -\frac{q^2}{4M^2}$$
where $G_M$ and $G_E$ are magnetic and electric proton formfactors.

A simple calculation gives the spin--independent and spin--dependent
parts of the well known Born cross--section in the form
\begin{equation}\label{20}
\frac{d\sigma^B}{dQ^2dy}=
\frac{4\pi\alpha^2(Q^2)}{Q^4y}[(1-y-xy\tau)F_2(x,Q^2)
+xy^2F_1(x,Q^2)]\ ,
\end{equation}
\begin{equation}\label{21}
\frac{d\sigma^B_{\parallel}}{dQ^2dy}= \frac{8\pi\alpha^2(Q^2)}{V^2y}\bigl
[\bigl(\tau-\frac{2-y}{2xy}\bigr)g_1(x,Q^2)+\frac{2\tau}{y}g_2(x,Q^2)\bigr]\ , \end{equation} \begin{equation}\label{22}
\frac{d\sigma^B_{\bot}}{dQ^2dy}= -
\frac{8\pi\alpha^2(Q^2)}{V^2y}\sqrt{\frac{M^2}{Q^2}(1-y-xy\tau)}\bigl[g_1(x,Q^2)
+\frac{2}{y}g_2(x,Q^2)\bigr]\ .
\end{equation}
Thus, within the leading accuracy, the radiatively corrected
cross--section
of the process (1) is defined by Eq.(2) (for its spin--independent part)
with (20) as a hard part of the cross--section and by Eq.(13) (for its
spin--dependent part) with (21) or (22) as a hard part.

It is useful to extract the first order correction to Born
approximation, as defined by master equation (2). For this purpose, we
can use
the iterative form of the photon structure function $D^{^{\gamma}}$ with
$L\rightarrow L-1$ and
$$\Delta\rightarrow\Delta_1 = \frac{2(\Delta\varepsilon)}{\sqrt{V}(1-xy)}
\sqrt{\tau+z_+}\ , \ z_+ =y(1-x)\ , \
\frac{2(\Delta\varepsilon)}{\sqrt{V}}\ll 1 $$
for $D(z_1,L)$ and
$$\Delta\rightarrow\Delta_2 = \frac{2(\Delta\varepsilon)}{\sqrt{V}(1-z_+)}
\sqrt{\tau+z_+}\ $$
for $D(z_2,L),$ where $(\Delta\varepsilon)$ is the minimal energy of hard
collinear photon in the special system $(\vec k_1-\vec k_2 +\vec p_1 =0).$
Straightforward calculations yield the following expression
\begin{equation}\label{23, first order leading}
\frac{d\sigma^{(1)}(k_1,k_2)}{dQ^2dy}=\frac{\alpha(L-1)}{2\pi}\Bigl\{
\frac{d\sigma^{(B)}(k_1,k_2)}{dQ^2dy}\Bigl[3+2\ln\frac{4(\Delta\varepsilon)^2
(z_++\tau)}{V(1-z_+)(1-xy)}\Bigr]+
\end{equation}
$$\int\limits_{z_{th}}^{z_+-\rho}dz\Bigl[\frac{1+z_1^2}{(1-xy)(1-z_1)}
\frac{d\sigma^{(B)}(z_1k_1,k_2)}{dQ_t^2dy_t} +
\frac{1+z_2^2}{(1-z_+)(1-z_2)}
\frac{d\sigma^{(B)}(k_1,k_2/z_2)}{dQ_s^2dy_s}\Bigr]\Bigr\}\ , $$
where
$$z=\frac{M_x^2-M^2}{V}\ , \ z_1=\frac{1-y+z}{1-xy}\ , \
z_2=\frac{1-z_+}{1-z}\ , \ \rho = \frac{2(\Delta\varepsilon)}{\sqrt{V}}
\sqrt{\tau+z_+}\ , $$
$$Q_t^2=-q_t^2=z_1Q^2\ , \ \ Q_s^2=-q_s^2={Q^2 \over z_2}\ , \ \ y_{t,s} =
1-\frac{1-y}{z_{1,2}} \ .$$

Similar equations can be derived for the
first order correction to
the spin--dependent part of the cross-section for both longitudinal and
transverse polarizations of the target proton.

\section{DIS cross--section beyond the leading
accuracy}
\label{sectNLODIS}

\hspace{0.7cm}

To go beyond the leading accuracy we have to improve the expressions for
hard parts of the cross--sections in master equations (2) and (13) to
include
effects caused by radiation of a hard noncollinear photon. (In
principle, we
can improve also the expression for $D$--function to take into account
collinear next--to--leading effects in the second order of perturbation
theory. The essential part of these effects is included in our
$D$--functions
due to replacement $L\rightarrow L-1.$ The rest can be written by using
the results of corresponding calculations for double photon emission
\cite{AAK,GKM,KM,MER}, pair production \cite{KMS,MERP,AKMP}, one
loop corrected Compton tensor \cite{AAK,GKM,KMFC} and virtual correction
\cite{Rem}. But here we restrict ourselves to $D$--functions given above
in Eqs.~(5), (6), (7) and (14)).

To compute the improved hard cross--section, one has to find the full
first
order RC to the cross--section of the process (1) and subtract from it (to
get rid of the double counting) its leading part that (for unpolarized
case)
is defined by Eq.~(23). Therefore, the improved hard part can be
written as
\begin{equation}\label{24, definition of hard}
\frac{d\sigma_{hard}}{dQ^2dy} = \frac{d\sigma^B}{dQ^2dy} +
\frac{d\sigma^{(S+V)}}{dQ^2dy}+\frac{d\sigma^H}{dQ^2dy}-
\frac{d\sigma^{(1)}}{dQ^2dy}\ ,
\end{equation}
where $d\sigma^{(S+V)}$ is a correction to the cross--section of the
process
(1) due to virtual and soft photon emission and $d\sigma^{H}$ is a
cross--section of the radiative process
\begin{equation}\label{25, radiative process}
e^-(k_1) + P(p_1) \rightarrow e^-(k_2) + \gamma(k) + X(p_x) \ .
\end{equation}

The virtual and soft corrections are factorized in the same form for both
polarized and unpolarized cases \cite{KMF} and can be written as
\begin{equation}\label{26, V+S}
\frac{d\sigma^{(S+V)}}{dQ^2dy} =
\frac{d\sigma^{B}}{dQ^2dy}\Bigl[1+\frac {\alpha}{2\pi}\Bigl(\delta
+ (L-1)\bigl(3+2\ln\frac{\rho^2}{(1-xy)(1-z_+)}\bigr)\Bigr)\Bigr]\
,
\end{equation}
$$\delta =
-1-\frac{\pi^2}{3}-2f\bigl(\frac{1-y-xy\tau}{(1-xy)(1-z_+)}\bigr) -
\ln^2\frac{1-xy}{1-z_+}\ , \ f(x) = \int\limits_0^x\frac{dt}{t}\ln(1-t)\
.$$

To calculate the cross--section of the radiative process (25), we use
the corresponding leptonic tensor in the form
\begin{equation}\label{27, radiative tensor}
L_{\mu\nu}^{\gamma} =
\frac{\alpha}{4\pi^2}(L_{\mu\nu}^{H(un)}+L_{\mu\nu}^H)\frac{d^3k}{\omega}
\ , \ L_{\mu\nu}^H=
2i\varepsilon_{\mu\nu\lambda\rho}q_{\lambda}(k_{1\rho}R_t +k_{2\rho}R_s)\
, \end{equation} $$R_t=
\frac{u+t}{st}-2m^2\bigl(\frac{1}{s^2}+\frac{1}{t^2}\bigr) \ , \ R_s=
\frac{u+s}{st}-2m^2\frac{s_t}{ut^2}\ , s_t = \frac{-u(u+Vy-Vz)}{u+V}\ , $$
where $\omega$ is the energy of radiated photon, $L_{\mu\nu}^{H(un)}$ is
the leptonic tensor for unpolarized particles, see Ref. \cite{KMFC}, and
we use the
following notation for kinematic invariants $$ \ s=2kk_2\ , \ \
t=-2kk_1\
,
\
\ q^2=u+s+t\ .	$$

The result for unpolarized case was derived in \cite{KMF}, and here we
rewrite it in terms of our standard notation
\begin{equation}\label{28, unpolarized part}
\frac{d\sigma_{hard}}{dQ^2dy} =
\frac{d\sigma^B}{dQ^2dy}\bigl(1+\frac {\alpha}{2\pi}\delta \bigr)
+ \frac{\alpha}{VQ^2}\int\limits_{z_{th}}^{z_+}dz
\Bigl\{\frac{1-r_1}{1-xy}\hat P_tN-\frac{1-r_2}{1-z_+}\hat P_sN +
\int\limits_{r_-}^{r_+}dr\frac{2W}{\sqrt{y^2+4xy\tau}} +
\end{equation}
$$+{\large{\it P}}\int\limits_{r_-}^{r_+}\frac{dr}{1-r}\Bigl[
\frac{1-\hat P_t}{|r-r_1|}\Bigl(\frac{(1+r^2)N}{1-xy}+(r_1-r)T_t\Bigr) -
\frac{1-\hat P_s}{|r-r_2|}\Bigl(\frac{(1+r^2)N}{1-z_+}+(r_2-r)T_s\Bigr)
\Bigr]\Bigr\}\frac{\alpha^2(rQ^2)}{r^2}\ , $$
where $r=-q^2/Q^2$ and the limits of the integration respect to $r$ are
$$r_{\pm}(z) =
\frac{1}{2xy(\tau+z_+)}[2xy(\tau+z)+(z_+-z)(y\pm\sqrt{y^2+4xy\tau})]\ . $$
Here we used the following notation
$$N=2F_1(x',r)+\frac{2x'}{rxy}\biggl(\frac{1-y}{xy}-\tau\biggr)F_2(x',r)
, \
W= 2F_1(x',r)-\frac{2x'\tau}{rxy}F_2(x',r)\ , $$
\begin{equation}\label{29,notations}
T_t= -\frac{2x'[1-r(1-y)]}{x^2y^2r}F_2(x',r)\ , \
T_s= -\frac{2x'(1-y-r)}{x^2y^2r}F_2(x',r)\ ,
\end{equation}
$$r_1=\frac{1-y+z}{1-xy}\ , \ \ r_2 = \frac{1-z}{1-z_+}\ , \ \ x'=\frac{
xyr}{xyr+z}\ .$$
The action of the operators $\hat P_t$ and $\hat P_s$ is defined as
follows
$$\hat P_tf(r,x') = f(r_1,x_t)\ , \ \ \hat P_sf(r,x') = f(r_2,x_s)\ , \ \
x_t=\frac{xyr_1}{xyr_1+z}\ , \ \ x_s=\frac{xyr_2}{xyr_2+z}\ . $$
Note that the quantity $r_1(r_2)$ coincides with $z_1(1/z_2)$ for
radiation of a single collinear photon.

The hard cross--section (29) has neither collinear nor infrared
singularities. The different terms on the right-hand side of Eq.~(29)
have
singularities at $r=r_1,\ r=r_2$ and $r=1.$ Singularities at first two
points are collinear and at third one is unphysical that arises at
integration. Collinear singularities vanish due to action of operators
$\hat P_t$ and $\hat P_s$ on the terms containing $N.$ The unphysical
singularity
cancels because in the limiting case $r\rightarrow 1$ we have
$$\frac{r_2-r}{|r_2-r|} =1\ , \ \frac{r_1-r}{|r_1-r|} =-1\ , T_t+T_s=0\
.$$

Let us consider the spin--dependent part of hard cross--section in more
details. The contraction of the spin--dependent parts of leptonic and
hadronic tensors can be written as
\begin{equation}\label{30, spin-dependent contraction}
L_{\mu\nu}^HH_{\mu\nu}^{^{\parallel,\bot}} =
-U^{^{\parallel,\bot}}C^{^{\parallel,\bot}}\frac{x'}{q^2}\ , \
U^{^{\parallel}}=1\ , \ U^{\bot} =
\frac{1}{V}\sqrt{\frac{M^2}{Q^2}(1-y-xy\tau)^{-1}}\ ,
\end{equation}
$$C^{^{\parallel,\bot}}=\widetilde 2W^{^{\parallel,\bot}}+
\Bigl[\frac{u^2+q_t^4}{t(q_t^2-u)}-\frac{2m^2}{ut^2}(u^2+q_t^2s_t)\Bigr]
\hat P_t\widetilde N_t^{^{\parallel,\bot}}+
\Bigl[\frac{u^2+q_s^4}{s(q_s^2-u)}-\frac{2m^2}{s^2}q_s^2\Bigr]
\hat P_s\widetilde N_s^{^{\parallel,\bot}} + $$
$$\frac{1-\hat P_t}{t}\frac{(u^2+q^4)\widetilde N_t^{^{\parallel,\bot}}+
2q^2(q_t^2-q^2)\widetilde T_t^{^{\parallel,\bot}}}{q^2-u}+
\frac{1-\hat P_s}{s}\frac{(u^2+q^4)\widetilde N_s^{^{\parallel,\bot}}+
2u(q_s^2-q^2)\widetilde T_s^{^{\parallel,\bot}}}{q^2-u}\ , $$
$$ q_t^2=\frac{uV(1-y+z)}{u+V}\ , \ \
q_s^2=\frac{uV(1-z)}{V(1-y)-u}\ . $$

For the case of longitudinal polarization of the target proton we have
\begin{equation}\label{31, longitudinal}
\widetilde W^{^{\parallel}}=4\tau[yx'Vg_2-(q^2+u)g_1]\ , \
\widetilde N_s^{^{\parallel}}=2\bigl(2q^2\tau+2V+\frac{q^2}{x'}\bigr)g_1-
8\tau x'Vg_2\ ,
\end{equation}
$$\widetilde
N_t^{^{\parallel}}=2\bigl[2u\tau+q^2(\frac{2V}{u}+\frac{1}{x'})\bigr]g_1-
8\tau x'Vg_2\ , \ \widetilde T_s^{^{\parallel}}= \frac{2uV(z-1)}{q_s^2}
(g_1-2\tau x'g_2)\ , $$
$$\widetilde T_t^{^{\parallel}}= 2(u+V)\bigl(\frac{q^2}{u}g_1-2\tau x'g_2
\bigr)\ .$$

The corresponding quantities for the case of transverse polarization of
the target proton read
\begin{equation}\label{32,perp}
\widetilde W^{^{\bot}}=2(2u\tau-Vy)[(q^2+u)g_1-x'yVg_2]\ ,
\end{equation}
$$\widetilde
N_s^{^{\bot}}=2\bigl[-uV-q^2(2u\tau+V(1-y)+\frac{u}{x'})\bigr]
\bigl(g_1-\frac{2x'V}{q^2}g_2\bigr)\ , $$
$$\widetilde
N_t^{^{\bot}}=2\bigl[-q^2V-u(2u\tau+V(1-y)+\frac{q^2}{x'})\bigr]
\bigl(g_1-\frac{2x'V}{u}g_2\bigr)\ , $$
$$\widetilde
T_t^{^{\bot}}=2(u+V)\bigl[-q^2g_1+\frac{x'V}{u}\bigl(q^2+u(1-y+
\frac{2u\tau}{V})\bigr)g_2\bigr] \ , $$
$$\widetilde
T_s^{^{\bot}}=2\frac{uV(1-z)}{q_s^2}\bigl[ug_1-\frac{x'V}{q^2}\bigl(u+q^2(1-y+
\frac{2u\tau}{V})\bigr)g_2\bigr] \ . $$

The action of operators $\hat P_t$ and $\hat P_s$ in the expressions for
$C^{^{\parallel,\bot}}$ can be understood if we write
$$r=\frac{q^2}{u} \ , \ \ r_1=\frac{q_t^2}{u} \ , \ \ r_2=\frac{q_s^2}{u}
\ , \ \ x'=\frac{q^2}{q^2-Vz}\ .$$

The cross--section of radiative process (25) can be written in terms of
the
quantities $C^{^{\parallel,\bot}}$ as follows
\begin{equation}\label{33, H-cross-section}
\frac{d\sigma^H_{\parallel,\bot}}{dQ^2dy}=-\frac{2\alpha}{V}U^{^{\parallel,\bot}}
C^{^{\parallel,\bot}}\frac{x'\alpha^2(-q^2)}{q^6}\omega d\omega
d\cos{\theta_k}\frac{d\phi}{2\pi}\ ,
\end{equation}
where $\theta_k$ and $\phi$ are polar and azimuth angles of photon in
the special system with $Z$-axis along the direction of the target
proton
3--momentum $\vec p_1$, provided $\vec k_1$ and $\vec k_2$ are
within
$XZ$ plane.

Integration of (33) with respect to photon variables can be done  in
full analogy with unpolarized case as described in \cite{KMF} (see
also \cite{AAMrec}). The result can be written in the following form
\begin{equation}\label{34, H-cross-section integrated}
\frac{d\sigma^H_{\parallel,\bot}}{dQ^2dy}=
-\frac{\alpha}{V}U^{^{\parallel,\bot}}\int\limits_{z_{th}}^{z_+-\rho}dz
\Bigl\{\frac{V}{u+V}\bigl[q_t^2-u-\frac{u^2+q_t^4}{q_t^2-u}(L-1)\bigr]\hat
P_t\widetilde N_t^{^{\parallel,\bot}} +
\frac{V}{V(1-y)-u}\bigl[q_s^2-u+
\end{equation}
$$\frac{u^2+q_s^4}{q_s^2-u}(L-1)\bigr]
\hat P_s\widetilde N_s^{^{\parallel,\bot}} +
P\int\limits_{q_-^2}^{q_+^2}\frac{dq^2}{q^2-u}\Bigl[\frac{V(1-\hat P_s)}
{(V(1-y)-u)|q^2-q_s^2|}\bigl((u^2+q^4)\widetilde N_s^{^{\parallel,\bot}}+
$$
$$2u(q_s^2-q^2)\widetilde T_s^{^{\parallel,\bot}}\bigr) -
\frac{V(1-\hat P_t)}
{(V+u)|q^2-q_t^2|}\bigl((u^2+q^4)\widetilde N_t^{^{\parallel,\bot}}
+2q^2(q_t^2-q^2)\widetilde T_t^{^{\parallel,\bot}}\bigr)\Bigr] + $$
$$\int\limits_{q_-^2}^{q_+^2}\frac{dq^2}{\sqrt{y^2+4xy\tau}}
\widetilde 2W^{^{\parallel,\bot}}\Bigr\}\frac{x'\alpha^2(-q^2)}{q^6}\ , \
q_{\pm}^2 = ur_{\mp} \ . $$

To derive the hard cross--section for the polarized case we have to add
(26)
and (34) without their leading contributions, which are proportional to
$L-1$ and  sum up to $d\sigma^{(1)}_{\parallel,\bot}/dQ^2dy.$
The result reads
\begin{equation}\label{35, DIS final}
\frac{d\sigma^{^{\parallel,\bot}}_{hard}}{dQ^2dy}=
\frac{d\sigma^B_{\parallel,\bot}}{dQ^2dy}\bigl(1+\frac{\alpha}{2\pi}\delta
\bigr)
+\frac{\alpha}{Q^4}\widetilde
{U}^{^{\parallel,\bot}}\int\limits_{z_{th}}^{z_+} dz
\Bigl\{\frac{1-r_1}{1-xy}\hat P_tN_t^{^{\parallel,\bot}}+
\frac{1-r_2}{1-z_+}\hat P_sN_s^{^{\parallel,\bot}}+
\end{equation}
$$+P\int\limits_{r_-}^{r_+}\frac{dr}{1-r}\Bigl[
\frac{1-\hat
P_s}{|r-r_2|(1-z_+)}\bigl((1+r^2)N_s^{^{\parallel,\bot}}+\frac{2(r_2-r)}{r_2}
T_s^{^{\parallel,\bot}}\bigr) -
\frac{1-\hat
P_t}{|r-r_1|}\bigl(\frac{(1+r^2)N_t^{^{\parallel,\bot}}}{1-xy}+$$
$$2r(r_1-r)
T_t^{^{\parallel,\bot}}\bigr)\Bigr] + \int\limits_{r_-}^{r_+}dr\frac
{2W^{^{\parallel,\bot}}}{\sqrt{y^2+4xy\tau}}\Bigr\}\frac{x'\alpha^2(Q^2r)}
{r^3}\ , $$
where
$$\widetilde U^{^{\parallel}}=1, \ \ \widetilde U^{^{\bot}} =
\sqrt{\frac{M^2}{Q^2}(1-y-xy\tau)^{-1}}\ ,  $$
$$W^{^{\parallel}} = 4y\tau W\ , \ \ W^{^{\bot}} = 2y^2(1+2x\tau)W\ , \ \
W=(1+r)xg_1+x'g_2\ , $$
$$N_t^{^{\parallel}}=2[2r-z-xy(r+2\tau)]g_1-8x'\tau g_2\ , \
N_s^{^{\parallel}}=2[2-z-xyr(1+2\tau)]g_1-8x'\tau g_2\ , $$
$$N_t^{^{\bot}} = 2[1-y-z+r-xy(r+2\tau)](xyg_1+2x'g_2)\ , \
N_s^{^{\bot}} = 2[1-y+\frac{1-z}{r}-xy(1+2\tau)](xyrg_1+2x'g_2)\ ,$$
$$T_t^{^{\parallel}}=2rg_1-4x'\tau g_2\ , \
T_s^{^{\parallel}}=2(z-1)(g_1-2x'\tau g_2)\ , $$
$$T_t^{^{\bot}}=2xyrg_1+2x'(1-y+r-2xy\tau)g_2\ ,
T_s^{^{\bot}}= 2(z-1)\big[xyg_1+x'(1-y+\frac{1}{r}-2xy\tau)g_2\bigr]\ .$$

The polarized hard cross--section defined by Eq.~(35) is free from
collinear singularities due to action of operators $1-\hat P_t$ and $1-\hat
P_s.$ The unphysical singularity at $r=1$ on the right-hand side of
Eq.~(35) cancels because in this limit
$$T_t{^{\parallel,\bot}}=\frac{1}{z-1}T_s{^{\parallel,\bot}}\ . $$

Note that radiation of photon at large angles by the initial
  and final electrons increases the region of variation for
  quantity $r$ in (35), because for collinear radiation
  $r_1<r<r_2$ and now $r_-<r_1$ and $r_+>r_2$. It may be
  important if the hadron structure functions are large in these
  additional regions.

\section{Hard cross--section for elastic hadronic events}
\label{sectNLOel}

\hspace{0.7cm}

To describe the hard cross--section for elastic hadronic events we use the
replacement defined by (19) in Eqs.~(28) and (35). For Born
cross-sections which enter in this equations, see Eqs.(21)--(23). The
function
$\delta(1-x')$ is used to do the integration with respect to
inelasticity
$z$
\begin{equation}\label{deltafun}
\int dz\delta(1-x') = xyr\ . \end{equation}
The final result for unpolarized case reads (we do not introduce special
notation for the elastic cross--section)
\begin{equation}\label{36,unpolarized elastic}
\frac{d\sigma_{hard}}{dQ^2dy}=\frac{d\sigma^B}{dQ^2dy}
\bigl(1+\frac{\alpha}{2\pi}
\delta\bigr)+ \frac{\alpha}{V^2}\Bigl\{\frac{1-r_1}{1-xy}\hat P_tN
 - \frac{1-r_2}{1-z_+}\hat P_sN +\int\limits_{r_-}^{r_+}dr\frac{2W}{\sqrt{
 y^2+4xy\tau}} +
\end{equation}
$$P\int\limits_{r_-}^{r_+}\frac{dr}{1-r}\Bigl[
\frac{1-\hat P_t}{|r-r_1|}\bigl(\frac{1+r^2}{1-xy}N+(r_1-r)T_t\bigr)-
\frac{1-\hat P_s}{|r-r_2|}\bigl(\frac{1+r^2}{1-z_+}N+(r_2-r)T_s\bigr)
\Bigr]\Bigr\}\frac{\alpha^2(Q^2r)}{r}\ , $$
where
$$N=G_M^2+\frac{2}{xyr}\bigl(\frac{1-y}{xy}-\tau\bigr)\frac{G_E^2+\lambda
G_M^2}{1+\lambda}\ , \ \
W=G_M^2-\frac{2\tau}{xyr}\frac{G_E^2+\lambda G_M^2}{1+\lambda}\ ,$$
$$T_t=-\frac{2}{x^2y^2r}[1-r(1-y)]\frac{G_E^2+\lambda G_M^2}{1+\lambda}\ ,
\ \ T_s=-\frac{2}{x^2y^2r}(1-r-y)\frac{G_E^2+\lambda G_M^2}{1+\lambda}\ .
$$
The Born cross-section on the right-hand side of Eq.(37) is defined as
\begin{equation}\label{37}
\frac{d\sigma^B}{dQ^2dy} = \frac{4\pi\alpha^2(Q^2)}{V^2}\Big[\frac{1}{2}
G_M^2+[1-y(1+\tau)]\frac{G_E^2+\lambda G_M^2}{y^2(1+\lambda)}\Bigr]
\delta\bigl(y-\frac{Q^2}{V}\bigr)\ .
\end{equation}
When writing this last equation we take into account that
$$\delta(1-x) = {y}\delta\bigl(y-\frac{Q^2}{V}\bigr)\ .$$

The spin--dependent hard cross--section for elastic hadronic
events can be written in the form very similar to (37)
\begin{equation}\label{38,polarized elastic}
\frac{d\sigma^{^{\parallel,\bot}}_{hard}}{dQ^2dy}=
\frac{d\sigma^B_{\parallel,\bot}}{dQ^2dy}\bigl(1+\frac{\alpha}{2\pi}
\delta\bigr)+
\frac{\alpha}{V}\widetilde
U^{\parallel,\bot}\Bigl\{\frac{1-r_1}{1-xy}\hat P_t
N_t^{^{\parallel,\bot}} + \frac{1-r_2}{1-z_+}\hat P_s
N_s^{^{\parallel,\bot}}
+\int\limits_{r_-}^{r_+}dr\frac{W^{^{\parallel,\bot}}}{\sqrt{ y^2+4xy\tau}}
+ \end{equation} $$P\int\limits_{r_-}^{r_+}\frac{dr}{1-r}\Bigl[
-\frac{1-\hat
P_t}{|r-r_1|}\bigl(\frac{1+r^2}{1-xy}N_t^{^{\parallel,\bot}}+2r(r_1-r)
T_t^{^{\parallel,\bot}}\bigr)+ $$
$$\frac{1-\hat
P_s}{|r-r_2|(1-z_+)}\bigl((1+r^2)N_s^{^{\parallel,\bot}}+
\frac{2(r_2-r)}{r_2}T_s^{^{\parallel,\bot}}\bigr)
\Bigr]\Bigr\}\frac{\alpha^2(Q^2r)}{(4M^2+Q^2r)r^2}\ , $$
where
$$W^{^{\parallel}}=4y\tau W\ , \ \ W^{^{\bot}} = 2y^2(1+2x\tau)W\ , \ \
W=r[x(1+r)-1]G_M^2+\bigl[r+\frac{4\tau}{y}(1+r)\bigr]G_MG_E\ , $$
$$ N_t^{^{\parallel}}=
r(2\tau+r)(2-xy)G_M^2+8\tau\bigl[r\bigl(\frac{1}{xy}-1\bigr)-\tau\bigr]
G_MG_E\ , $$
$$N_s^{^{\parallel}}=
r(2\tau+1)(2-xyr)G_M^2+8\tau\bigl[\frac{1}{xy}-r(1+\tau)\bigr]
G_MG_E\ , $$
$$N_t^{^{\bot}}=[1-y+r-xy(r+2\tau)][-r(2-xy)G_M^2+2(r+2\tau)G_MG_E]\ , $$
$$N_s^{^{\bot}}=\bigl[1-y+\frac{1}{r}-xy(1+2\tau)\bigr][-r(2-xyr)G_M^2
+2r(1+2\tau)G_MG_E]\ , $$
$$T_t^{^{\parallel}}=r\bigl[(r+2\tau)G_M^2+2\tau\bigl(\frac{2}{xy}-1\bigr)
G_MG_E\bigr]
, \ \ T_s^{^{\parallel}}=-r(1+2\tau)G_M^2-2\tau\bigl(\frac{2}{xy}-
r\bigr)G_MG_E\ , $$
$$T_t^{^{\bot}}=r\bigl\{-[r(1-xy)+1-y-2xy\tau)]G_M^2+[1-y-2xy\tau+
r+4\tau]G_MG_E \bigr\} \ , $$
$$T_s^{^{\bot}}=r\bigl[\frac{1}{r}-xy(1+2\tau)+1-y\bigr]G_M^2-\bigl[2\tau(
2-xyr)+1+r(1-y)\bigr]G_MG_E\ .$$ Note that the argument of
electromagnetic formfactors in Eqs.(37) and (39) is $-Q^2r.$

The Born cross--sections on the right-hand side of Eq~(39) have the
following form
\begin{equation}\label{39,Born par}
\frac{d\sigma^B_{\parallel}}{dQ^2dy}=
\frac{4\pi\alpha^2(Q^2)}{V(4M^2+Q^2)}\Bigl[4\tau\bigl(1+\tau-\frac{1}{y}
\bigr)G_MG_E -(1+2\tau)\bigl(1-\frac{y}{2}\bigr)G_M^2\Bigr]\delta\bigl
(y-\frac{Q^2}{V}\bigr) \ ,
\end{equation}
for the longitudinal polarization of the target proton and
\begin{equation}\label{40,Born perp}
\frac{d\sigma^B_{\bot}}{dQ^2dy}=\frac{8\pi\alpha^2(Q^2)}{V(4M^2+Q^2)}
\sqrt{\frac{M^2}{Q^2}[1-y(1+\tau)]}\Bigl[\bigl(1-\frac{y}{2}\bigr)G_M^2-
(1+2\tau)G_MG_E\Bigr]\delta\bigl(y-\frac{Q^2}{V}\bigr) \ ,
\end{equation}
for the transverse one. The argument of formfactors in (40),
(41) is $-Q^2.$

\section{Numerical estimations}
\label{sectNum}

The formulae obtained in the last section include some operators which
emphasize the physical meaning of made transformations. However
they
are not convenient to numerical analysis. Here we present a unified
version of the formulae without any operators. For example,
the symbol $P$ is explicitly treated as
$$
P\int\limits_{r_-}^{r_+}\frac{d\,r}{1-r}F(r)=
\int\limits_{r_-}^{r_+}\frac{d\,r}{1-r}\bigl(F(r)-F(1)\bigr)
+F(1)\log{1-r_- \over r_+-1}
$$

So the formula reads
\begin{eqnarray}\label{unified}
\frac{d\sigma_{hard}^i}{dQ^2dy}&=&
\frac{d\sigma^B_i}{dQ^2dy}\bigl(1+\frac
{\alpha}{2\pi}\delta\bigr) +
{\alpha}U_i\int\limits_{z_{th}}^{z_+}dz
\biggl(
\Bigl\{ L_1^iN_i(r_1) + L_2^iN_i(r_2) \Bigr\}
+
\int\limits_{r_-}^{r_+}dr
\biggl\{W_i + T_i+
\\ && \!\!\!\!
+\frac{1}{1-r}\biggl[
N_i(r_1)-N_i(r_2)
\nonumber
+\ \frac{1-r_1}{|r-r_1|}  \Bigl[ N_i(r)-N_i(r_1) \Bigr]
+ \frac{1-r_2}{|r-r_2|}  \Bigl[ N_i(r)-N_i(r_2) \Bigr]
\biggr]
\biggr\}
\biggr)
\nonumber
\end{eqnarray}
where
$$
L^i_{1,2}=\mp b_i {(1-r_{1,2})^2 \over 1+r_{1,2}^2} \mp \log{1-r_-
\over r_+-1}, \quad b_u=-1, \quad b_{l,t}=1.
$$
The index $i$ runs over all polarization states ($i=u,l,t$). The
functions
$N_i(r)$ and $T_i$ read
$$
N_i(r)={1+r^2 \over z_+-z} N_i {x'\alpha^2 \over r^3},
\qquad
 T_i=\left\{
\begin{array}{ll}
\displaystyle	 \pm {T_{i1} \over 1-r} {x'\alpha^2\over r^3}  &
\displaystyle r<r_1;
\;\;\;\; r>r_2 \\
 \displaystyle	   T_{i2} {x'\alpha^2\over r^3}    & \displaystyle
r_1<r<r_2
\end{array}
\right.
$$
the pole $r=1$ can be reached only in the region $r_1<r<r_2$, so there
is no singularity in terms with $T_{i1}$. For $T_{i2}$ this pole is
explicitly canceled:
$$
T_{u2}={2(2-y)F_2 \over x^2y^2 }, \;\;
T_{l2}= -4(1+r)g_1 +8x'\tau g_2, \;\;
T_{t2}=
-4(1+r)xyg_1
-4x'(r+{1 \over r} +2-y - 2xy\tau )g_2.
$$
For unpolarized case $N_u=rN/x'$ with $n$ from (\ref{28, unpolarized
part}). For other cases they are
\begin{eqnarray}
N_l&=& 2\biggl[-1-r+\frac{y(1+2x\tau)[1-z+r(1-xy)]}{2-y}\biggr] g_1
+8x'\tau g_2
\
,
\nonumber \\
N_t&=& -\frac{4[1-z+r(1-xy)]}{r(2-y)}\biggl[xyr(1-y-xy\tau )g_1 +
x'(1-y+z+r(1-y+xy))g_2\biggr] +
\nonumber \\ &&
+4x'y(1+2x\tau)g_2 \ , \nonumber \\
\nonumber \\
T_{u1}&=&-{2(1+r)F_2 \over x^2y }
\nonumber \\
T_{l1}&=& 4\frac{y(1+r^2)(1+2x\tau)}{2-y}g_1 +8x'(1+r)\tau g_2 \ ,
\nonumber
\\ T_{t1}&=& 4\biggl
\{\frac{1+r^2}{2-y}\biggl[
-2xy(1-y-xy\tau)g_1
+(y-2z+yr(1-2x))\frac{x'}{r}g_2
\biggr] \nonumber
\\&&
+ x'y(1+2x\tau)(1+r)g_2\biggr \}
\nonumber \\ &&
\nonumber \end{eqnarray}
and $$ W_u={2W \over \sqrt{y^2+4xy\tau}}
{\alpha^2 \over r^2}, \qquad W_{l,t}={2W^{\parallel,\bot} \over
\sqrt{y^2+4xy\tau}} {x\alpha^2 \over r^3} $$ $$ U_u={1 \over V Q^2} \qquad
U_{l,t}={\widetilde U^{\parallel,\bot} \over Q^4}
$$

For elastic case the same formulae can be kept. Only the
formulae (\ref{19, elastic limit}) and (\ref{deltafun}) are needed here.
So for elastic case one
has to substitute
$$
\int dz \rightarrow xyr,
$$
setting $x'=1$, $z=0$ and structure functions in accordance with
(\ref{19, elastic limit}).

%

\begin{figure}[t!]\centering
\unitlength 0.6mm
\parbox{.48\textwidth}{\centering
\begin{picture}(100,100)
\put(-23,-10){
\epsfxsize=9cm
\epsfysize=9cm
\epsfbox{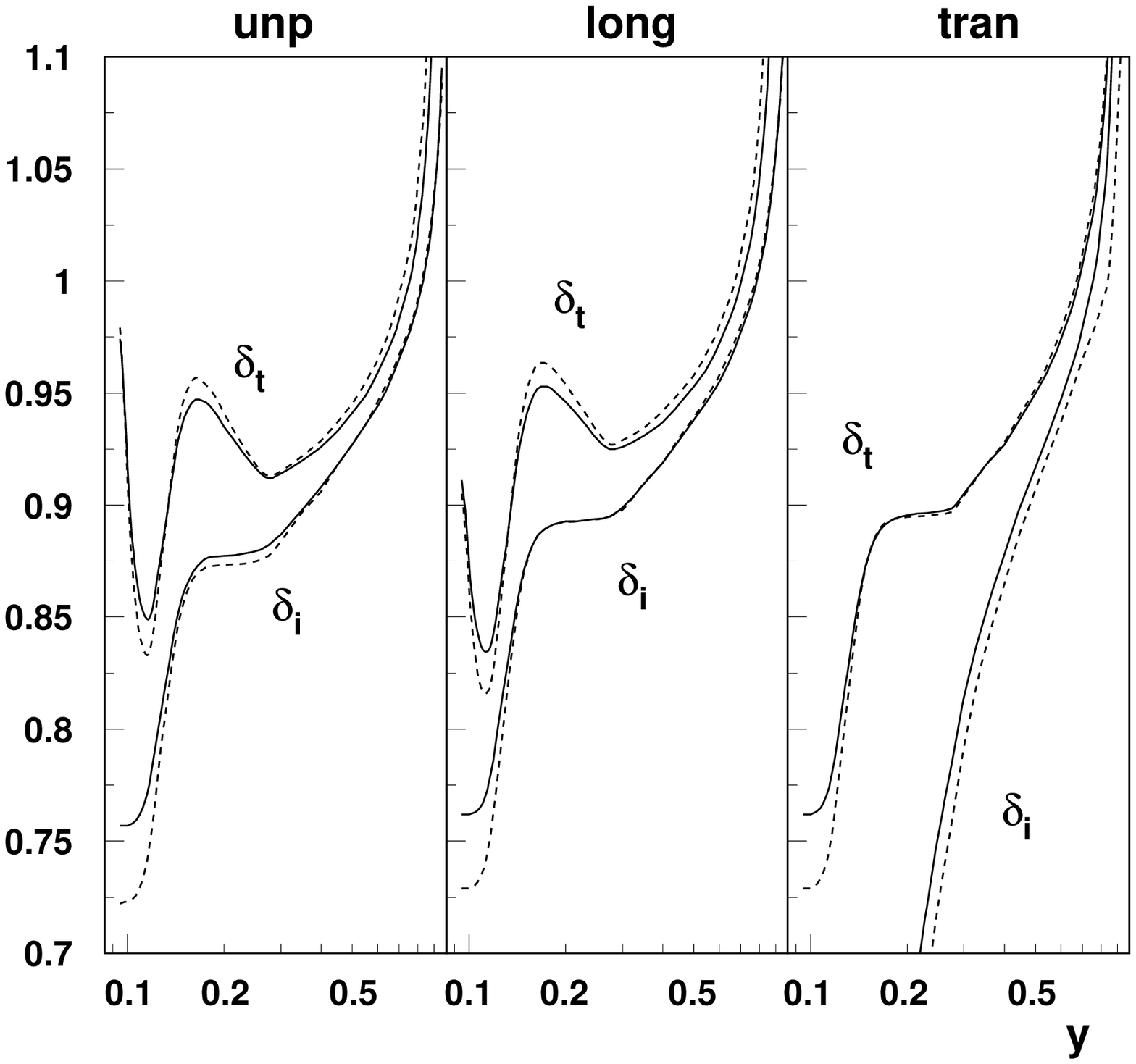}
}
\end{picture}
\caption{\label{Fig1}
Radiative correction to unpolarized and polarized (both longitudinal
and transverse) parts of cross section for kinematics close to JLab
experiments, $V$=10 GeV$^2$,
$x$=0.5.
} } \hfill
\parbox{.48\textwidth}{\centering
\begin{picture}(100,100)
\put(-23,-10){
\epsfxsize=9cm
\epsfysize=9cm
\epsfbox{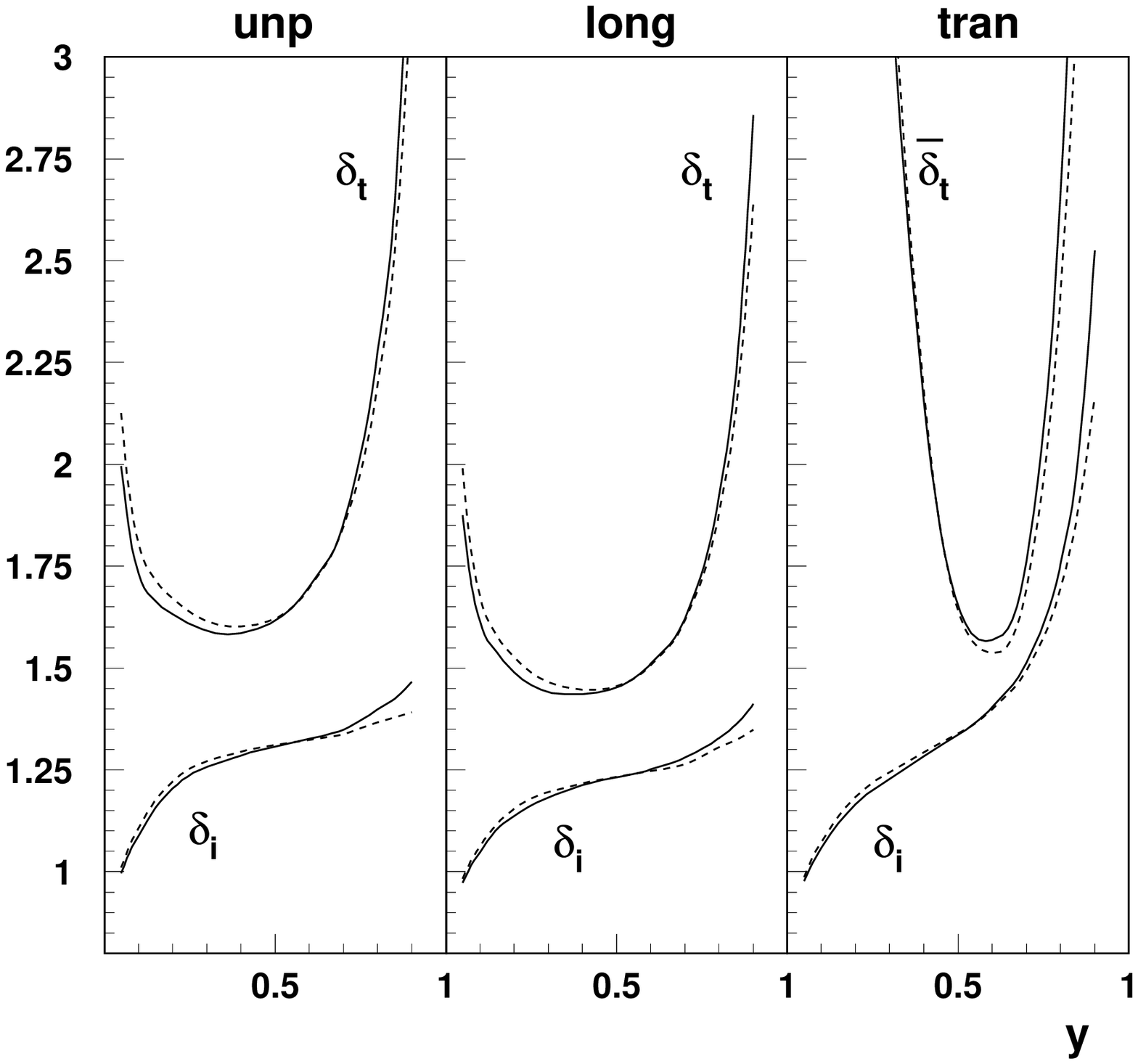}
}
\end{picture}
\caption{\label{Fig2}
Radiative correction to unpolarized and polarized (both longitudinal
and transverse) parts of cross section for kinematics close to HERMES
experiment, $V$=50
GeV$^2$,
$x$=0.1.
$\bar \delta _t=-\delta_t$
}
}
\end{figure}

It is believed that the formulae obtained within the presented formalism
are
not convenient for numerical analysis. There are two
reasons for such an opinion. First, the electron structure function in
the form
(\ref{5},\ref{6,pair contribution into EST}) has very sharp peak for $z$
going to unity. Secondly,
due to appearing of absolute values in denominators, the integrand
cannot be a continuous
function of the integration variables. It produces obstacles for
numerical
analysis if it is carried out in traditional style based on adaptive
methods of numerical integration, which is used in such programs as
TERAD/HECTOR \cite{HECTOR}
or POLRAD \cite{POLRAD}. However, it is possible to perform numerical
analysis if instead of adaptive integration we use Monte Carlo
integration while extracting the regions with sharp peaks into separate
integration subregions. Based on these ideas we developed Fortran
code ESFRAD\footnote{Electron Structure a Function method
for RADiative corrections} which allows one to perform the
numerical analysis without any serious difficulties.

We considered two radiative processes. In the first case, continuum of
hadrons is produced, while in the second case the proton remains in
the ground state.
Both of the considered effects contribute to the
experimentally observed cross section\footnote{Here and below we mean
double differential cross section $\sigma=d\sigma/dydQ^2$} of DIS. They
are usually called radiative tails from the continuous spectrum and the
elastic
peak or simply inelastic and elastic radiative tails. Below we study the
contributions of the tails numerically within kinematical conditions of
the current experiments on DIS.

\begin{figure}[ht]
\unitlength 1mm
\begin{center}
\begin{picture}(160,80)
\put(0,-10){
\epsfxsize=16cm
\epsfysize=10cm
\epsfbox{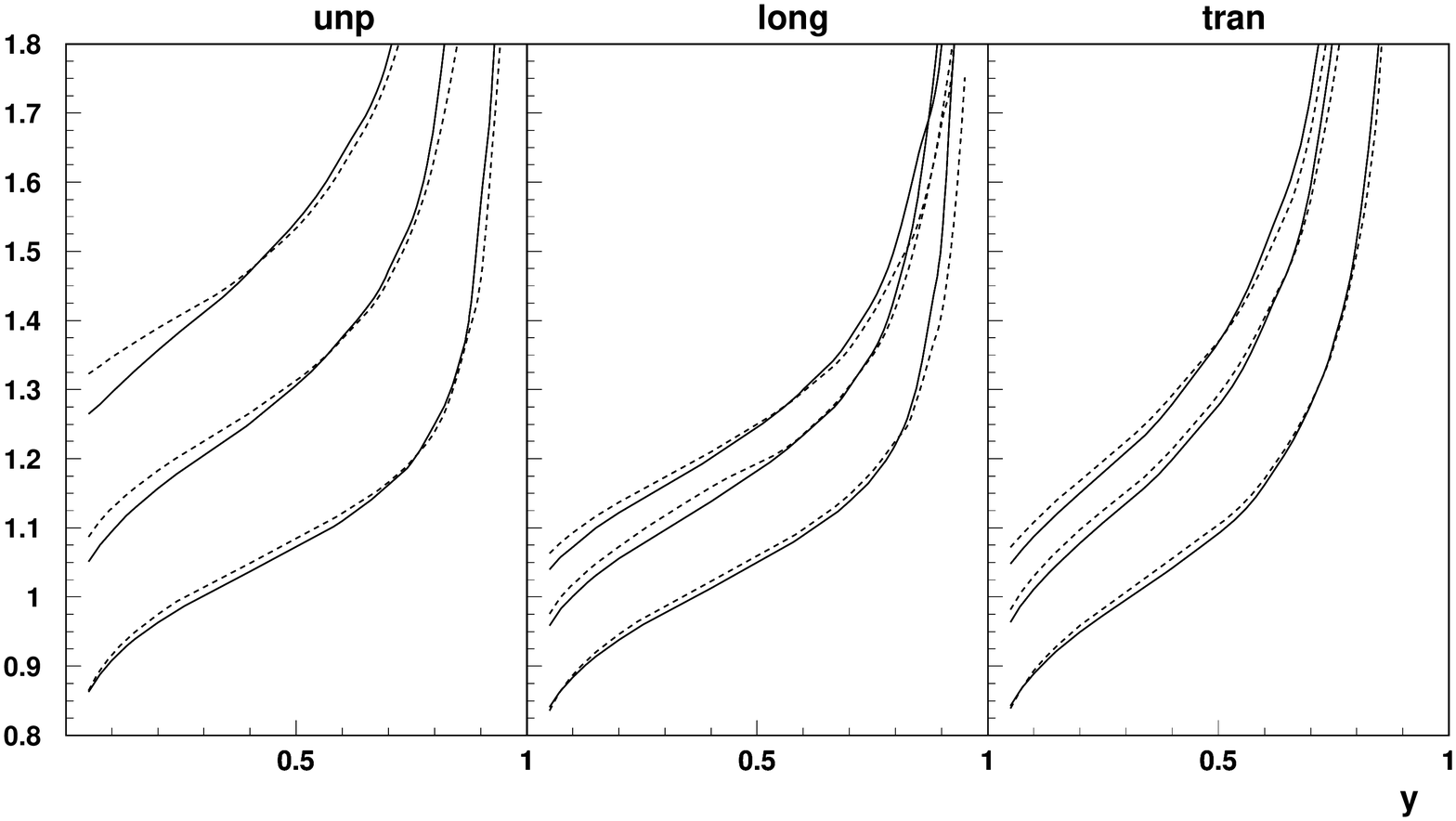}
}
\end{picture}
\end{center}
\caption{\label{Fig3}
One loop and total radiative correction (dashed and solid lines) for collider
kinematics (HERA); $V$=10$^5$GeV$^2$. Lines from top to bottom correspond
to
different values of $x$=0.001, 0.01 and 0.1}
\end{figure}

We take three typical values of $V$ equal to 10, 50 and 10000
GeV$^2$. They correspond to JLab, HERMES and HERA measurements.
Figures \ref{Fig1}, \ref{Fig2} and \ref{Fig3} give the radiative
correction factor for all polarization states (unpolarized, longitudinal
or transverse)
\begin{equation}
\delta_{i,t}={\sigma^{obs} \over \sigma^B }.
\end{equation}
The observed double differential cross section is given by
the master formulae
(\ref{2,master formula},\ref{13,master formula for
polarized case}), while the Born cross
section is calculated as
(\ref{20},\ref{21},\ref{22}). Both elastic and inelastic contributions
have to be taken for $\sigma_{hard}$. In this case we obtain the total
RC factor ($\delta_{t}$).
Subscripts ${i,t}$ correspond to the cases when elastic radiative
tail
is included into total correction ($\delta_{t}$) or inelastic radiative
tail contributes only ($\delta_{i}$). The elastic radiative tail
optionally may not be included because
sometimes there exist experimental methods to separate this contribution.
We note that for HERA kinematics we do not include it because it is
usually separated experimentally. Also we can extract a one-loop
contribution in order to study the
effect of higher order correction. The observed cross section in this
case is defined by the sum of the cross sections defined in eqs.
(\ref{23, first order leading})
and (\ref{unified}). We note that it can provide an additional cross check
by comparison with POLRAD.

We use rather simple models for spin-averaged and
spin-dependent structure functions. It allows us not to mix pure radiative
effects, which are of interest, with influence of hadron structure
functions.
Specifically, we use the so-called D8 model for spin-average SF
\cite{D8}
(see
also discussion in\cite{POLRAD}), and $A_1(x)=x^{0.725}$ suggested in
\cite{naga}; $g_2=0$ (for definition $A_1(x)$ see below).

From these plots we can see that the total radiative correction is
basically
defined by one-loop correction with some important effect around
kinematical borders. The sign and value of the higher order effects is in
agreement with	leading log estimations and calculations of correction
to elastic radiative tail in refs.\cite{ABSh,AKSh}. Two regions require
special consideration, namely, the region of higher $y$ for HERMES and
JLab kinematics
and   the region near pion threshold at JLab.

\begin{figure}[t!]\centering
\unitlength 0.6mm
\parbox{.48\textwidth}{\centering
\begin{picture}(100,100)
\put(-23,-10){
\epsfxsize=9cm
\epsfysize=9cm
\epsfbox{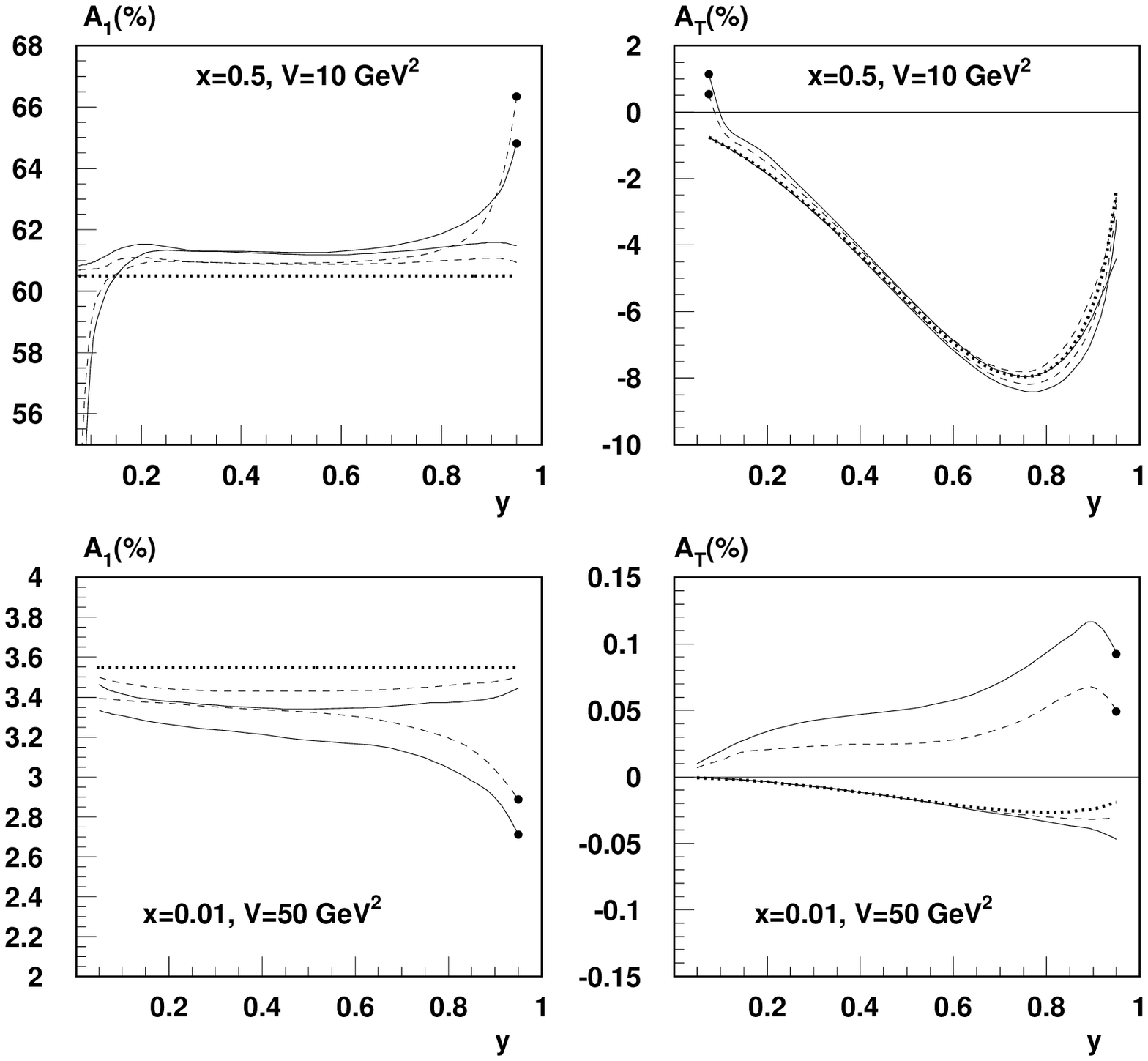}
}
\end{picture}
\caption{\label{Fig4}
Radiative correction to asymmetries for HERMES (lower plots) and JLab
(upper plots) kinematics. Dotted line shows Born asymmetry. Full and
dashed line correspond to total and one-loop contributions. Asymmetries
with taking into account of elastic contribution are marked by dots in the
end
 } } \hfill
\parbox{.48\textwidth}{\centering
\begin{picture}(100,100)
\put(-23,-10){
\epsfxsize=9cm
\epsfysize=9cm
\epsfbox{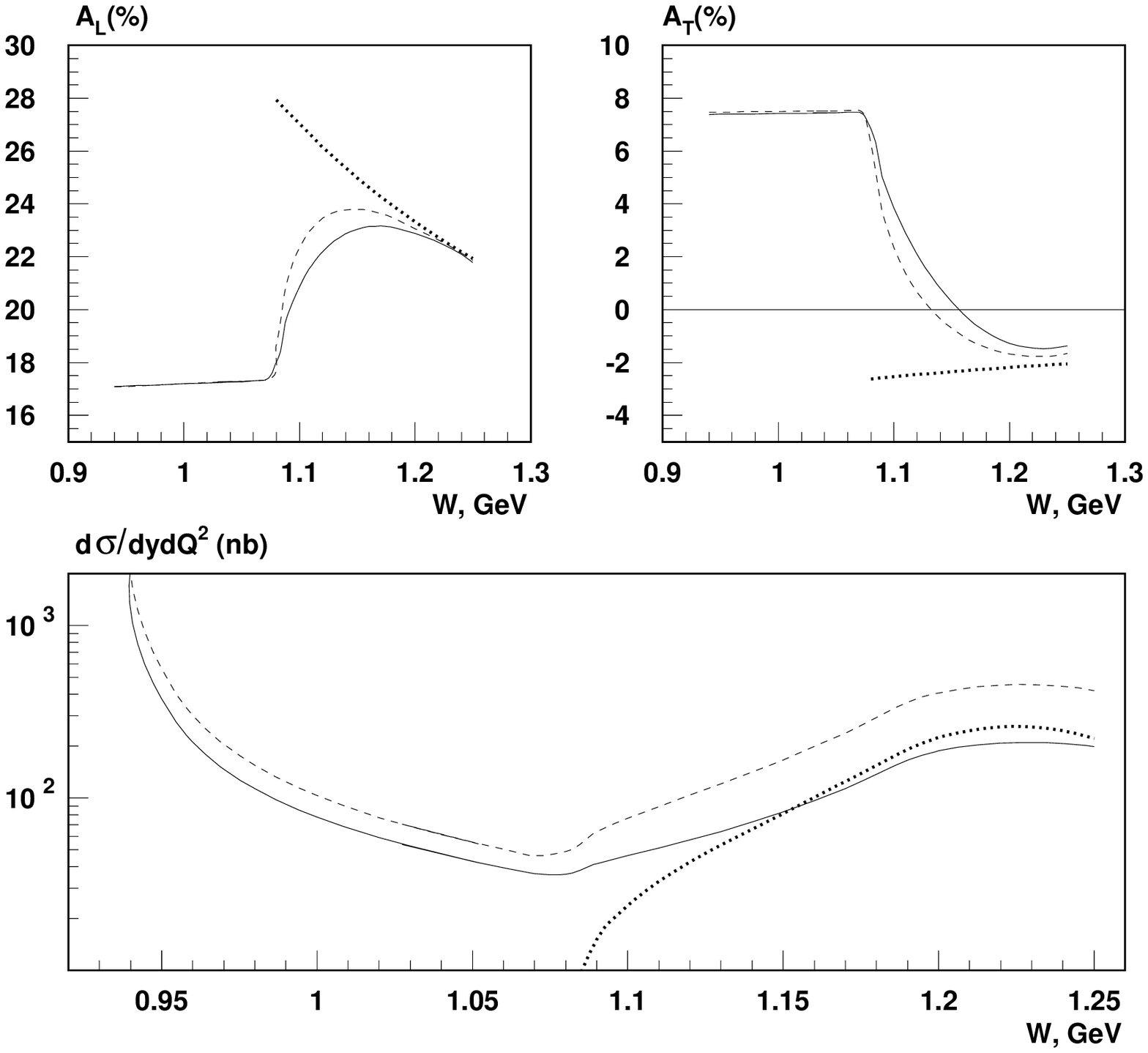}
}
\end{picture}
\caption{\label{Fig5}
The cross section (lower plot) and polarization asymmetries
 (both
longitudinal
and transverse) asymmetries for JLab kinematics ($Q^2$=1 GeV$^2$) near pion
threshold.
 Dotted line shows Born cross section and asymmetry. Full and
dashed line correspond to total and one-loop contributions.
}
}
\end{figure}

Let us define the polarization asymmetries as usual
\begin{equation}
A_L={\sigma_{\|} \over \sigma}, \qquad
A_T={\sigma_{\perp} \over \sigma}.
\end{equation}
Also we can define spin asymmetry $A_1$ which (for chosen model where
$g_2=0$) is simply related with $A_L=DA_1$, where $D$ is kinematical
depolarization factor dependent on the ratio
$R$ of longitudinal and transverse photoabsorption
cross sections
$$
D={y(2-y)(1+\gamma^2y/2)\over
y^2(1+\gamma^2)+2(1-y-\gamma^2y^2/4)(1+R)},
\qquad
R={\sigma_L \over \sigma_T}={M(Q^2+\nu^2)\over Q^2\nu} {F_2\over F_1}-1,
.
$$
where $\nu=yV/2M$ and $\gamma^2=Q^2/\nu^2$.
For
fixed $x$ $A_1$ is a constant within our
model, so it is very convenient for graphical presentation and analysis of
different radiative effects. Figure \ref{Fig4} gives asymmetries $A_1$ and
$A_T$ for kinematics of HERMES and JLab up to $y=0.95$. Influence of
higher order and elastic radiative effects can be seen. Figure \ref{Fig5}
gives total corrections to cross sections and asymmetries for threshold
region of JLab.

\section{Conclusion}
\label{sectConc}


In this paper we consider model-independent QED radiative
correction to the polarized DIS and elastic electron--proton
scattering. Together with analytical expression for RC, we
give its numerical values for different experimental
situations.

   Our analytical calculations are based on the electron
structure function method which allows to write both the
spin-independent and spin-dependent parts of the cross
section with accounting of RC to the leptonic part of
interaction in the form of well known Drell-Yan
representation. The corresponding RC includes explicitly the
first order correction as well as the leading-log contribution
in all orders of perturbation theory and the main part of the
second order next-to-leading-log one. Moreover, any
model-dependent RC to the hadronic part of interaction can be
included in our analytical result by inserting it as an additive
part of the hard cross section under integral sign in master
formulae (2) and (13).

   To derive RC, we take into account radiation of photons and
$e^+e^-$ pairs in collinear kinematics which produces a
large logarithm $L$ in the radiation probability (in $D$-functions) and
radiation of one  non-collinear photon that enlarges the limits
for variation of the hadron structure function arguments. It may
be important that these functions are sharp enough. In this case
the loss in radiation probability (the loss of $L$) can be
compensated by the increase in the value of the hard cross
section.

Note that we extracted the explicit formulae for the first order both
with LO and NLO levels. We found analytical agreement between these
results for the one-loop correction with the ones known earlier from
paper
\cite{ASh}, that provides
the most important test of total
correction.

On the basis of the analytical results, we constructed Fortran code
ESFRAD,\footnote{Fortran code ESFRAD is available at
http://www.jlab.org/$\sim$aku/RC}. Due to several known reasons
discussed in Section \ref{sectNum}
 the
results obtained by electron structure method is usually not so
convenient
for precise numerical analysis. However, we believe that found numerical
procedure based on Monte Carlo integration allows us to overcome
the obstacles.

Using the developed code we performed numerical analysis for kinematical
conditions of current and future polarization experiments.
We found two kinematical regions where the higher order radiative
correction
can be important. These are the traditional region of high $y$ and the
region
around the pion threshold. We gave detailed analysis of the effects
within
these regions and presented numerical results within one of the simplest
possibility for modeling DIS structure functions. Model dependence of
the result is surely an important question requiring a separate
investigation for specific application within experimental data
analysis.

\section*{Acknowledgements}
We thank our colleagues at Jefferson Lab for useful discussions. We thank the
US Department of Energy for support under contract DE-AC05-84ER40150. Work of
NM was in addition supported by Rutgers University through NSF grant PHY
9803860 and by Ukrainian DFFD. AA acknowledges additional support through NSF
grant PHY-0098642.

\end{document}